\documentclass{aa}

\usepackage{graphicx}
\usepackage[varg]{txfonts}
\usepackage{natbib}
\bibpunct{(}{)}{;}{a}{}{,}
\usepackage{array}
\usepackage{xspace}
\usepackage{lscape}
\usepackage{url}
\usepackage{arydshln}
\usepackage[hyperfootnotes=false, linktocpage=true, breaklinks=true, colorlinks=true, linkcolor=blue, citecolor=blue, urlcolor=blue]{hyperref}
\usepackage[all]{hypcap}
\usepackage{longtable}
\usepackage{afterpage}
\usepackage[dvipsnames]{xcolor}
\usepackage{xfrac}

\newcommand{\FeKa}{Fe K\ensuremath{\alpha}\xspace}

\newcommand{\fexxv}{\ion{Fe}{xxv}\xspace}
\newcommand{\fexxvi}{\ion{Fe}{xxvi}\xspace}
\newcommand{\kms}{\ensuremath{\mathrm{km\ s^{-1}}}\xspace}
\newcommand{\NH}{\ensuremath{N_{\mathrm{H}}}\xspace}
\newcommand{\vout}{\ensuremath{v_{\mathrm{out}}}\xspace}
\newcommand{\sigv}{\ensuremath{\sigma_{v}}\xspace}

\newcommand{\slab}{\xspace{\tt slab}\xspace}
\newcommand{\xabs}{\xspace{\tt xabs}\xspace}
\newcommand{\hot}{\xspace{\tt hot}\xspace}

\newcommand{\nustar}{{\it NuSTAR}\xspace}
\newcommand{\xmm}{{\it XMM-Newton}\xspace}
\newcommand{\xrism}{{XRISM}\xspace}
\newcommand{\chandra}{{\it Chandra}\xspace}

\newcommand{\hst}{{HST}\xspace}
\newcommand{\nicer}{{\it NICER}\xspace}

\newcommand{\ergs}{{\ensuremath{\rm{erg\ s}^{-1}}}\xspace}
\newcommand{\cm}{{\ensuremath{\rm{cm}^{-2}}}\xspace}
\newcommand{\spex}{\xspace{\tt SPEX}\xspace}
\newcommand{\pion}{\xspace{\tt pion}\xspace}

\newcommand{\logxi}{\ensuremath{{\log \xi}}\xspace}
\newcommand{\MBH}{\ensuremath{{M_{\rm BH}}}\xspace}
\newcommand{\lion}{\ensuremath{{L_{\rm ion}}}\xspace}
\newcommand{\rmin}{\ensuremath{{r_{\rm min}}}\xspace}
\newcommand{\rmax}{\ensuremath{{r_{\rm max}}}\xspace}

\newcommand{\ngc}{{NGC~3783}\xspace}

\mathchardef\mhyphen="2D

\begin{document}

\title{Delving into the depths of NGC 3783 with XRISM}
\subtitle{V. Broad-band modeling of ionized outflows}

\author{
Keqin Zhao \inst{1}
\and
Jelle S. Kaastra \inst{1,2}
\and
Liyi Gu \inst{1,2}
\and
Missagh Mehdipour \inst{3}
\and
Megan E. Eckart \inst{4}
\and
Keigo Fukumura \inst{5}
\and
Matteo Guainazzi \inst{7}
\and
Chen Li \inst{1,2}
\and
Christos Panagiotou \inst{6}
\and
Matilde Signorini \inst{7,8}}

\institute{
% 1
Leiden Observatory, Leiden University, PO Box 9513, 2300 RA Leiden, the Netherlands \\ \email{kzhao@strw.leidenuniv.nl}
\and 
% 2
SRON Netherlands Institute for Space Research, Niels Bohrweg 4, 2333 CA Leiden, the Netherlands
\and
%3
Department of Astronomy, University of Michigan, MI 48109, USA
\and
% 4
Lawrence Livermore National Laboratory, Livermore, CA 94550, USA
\and
% 5
Department of Physics and Astronomy, James Madison University, Harrisonburg, VA 22807, USA
\and
% 6
MIT Kavli Institute for Astrophysics and Space Research, Massachusetts Institute of Technology, Cambridge, MA 02139, USA
\and
% 7
ESA European Space Research and Technology Centre (ESTEC), Keplerlaan 1, 2201 AZ, Noordwĳk, the Netherlands
% 8
\and
Osservatorio Astrofisico di Arcetri, Largo Enrico Fermi 5, I-50125 Florence, Italy
}
\date{Received January 23, 2026; accepted April 13, 2026}

%###################################################
%###################################################
\abstract
{The Seyfert 1 galaxy NGC 3783 hosts a multiphase warm absorber (WA) that has been extensively studied in the X-ray band. High-resolution spectra from 2000--2001 revealed a complex outflow with multiple ionization and velocity components. Two decades later, new \xmm and \xrism observations allow us to investigate the long-term evolution of these outflows. We perform joint spectral modeling of the  \xmm/RGS and \xrism/Resolve time-averaged spectra using the \pion photoionization code within \spex. We derive the ionization parameter, column density, turbulent velocity, and outflow velocity for each absorption component, and investigate their thermal stability and Absorption Measure Distribution (AMD) to characterize the physical and dynamical properties of the WA in NGC 3783 in 2024. We compare these results with the 2000--2001 epoch to assess long-term variability, stability, and possible changes in the absorber population. We identify eight WA components spanning $\logxi = 1.08$--3.38 and outflow velocities of 480--1230 km s$^{-1}$. The ranges of column densities and turbulent velocities remain broadly consistent with the WAs from 2000--2001, but the earlier data contained more low-ionization, high-velocity components. The total column density in 2024 is 1.5 times larger than in 2000--2001, requiring replenishment by fresh material. The dominant Unresolved Transition Array (UTA) absorber (Comp. B3) has increased its column density by a factor of three while maintaining a similar ionization parameter. The WAs in NGC 3783 have undergone significant structural and dynamical evolution over the past 24 years.}
\keywords{X-rays: galaxies -- galaxies: active -- galaxies: Seyfert -- galaxies: individual: NGC 3783 -- techniques: spectroscopic}
\authorrunning{K. Zhao et al.}
\titlerunning{Delving into the depths of NGC 3783 with XRISM. V.}
\maketitle
\nolinenumbers

%#######################################################
%#######################################################
\section{Introduction}
\label{sect_intro}

Active galactic nuclei (AGN) outflows represent one of the most important mechanisms for AGN feedback, fundamentally influencing galaxy evolution and the regulation of star formation across cosmic time \citep{Fabi12,King15,Gasp17}. Among Seyfert 1 galaxies, NGC 3783 stands out as an exceptional laboratory for studying the complex physics of photoionized outflows, offering unique insights into the nature of AGN-driven winds and their impact on the surrounding environment \citep{kasp01,Beh03}.

NGC 3783 has been selected as a prime target for outflow investigations due to several favourable observational characteristics \citep{Netz2003}. As a bright Seyfert 1 galaxy (z = 0.0097), it provides sufficient X-ray flux for high-resolution spectroscopic studies while maintaining relatively simple geometric constraints. The nucleus is viewed at a relatively unobscured inclination typical of Seyfert 1 systems, ensuring that the spectral decomposition is physically unambiguous \citep{Blu02}. This geometry provides a direct line of sight to the central X-ray source and the intervening ionized gas, thereby simplifying the interpretation of absorption features arising from the outflow \citep{Tur08}.

The source exhibits well-defined absorption features across multiple ionization states, making it possible to trace the structure of the outflowing material \citep{Beh03}. NGC 3783 represents the best case for studying the nature of photoionized outflows, as various types of photoionized outflows are observed from this AGN. These include the classical multi-phase warm absorber detected across a wide range of ionization states in the soft X-ray and UV bands \citep{Gabe03}, a variable obscurer \citep{Mehd17}, an ultrafast outflow associated with a possible disk wind \citep{Mehdi25,GU25}, and photoionized emission from both the broad and narrow line regions \citep{Mao19}. Furthermore, the source's moderate redshift places key diagnostic lines within the optimal energy ranges of current X-ray observatories, enabling detailed kinematic and ionization analysis of the outflowing gas \citep{kasp01,Mao19}. The foundation for our current understanding of NGC 3783's outflows was established during the 2000--2001 multiwavelength monitoring campaign \citep{kasp01, Beh03}. It established the multi-component nature of the absorption system and provided the first detailed kinematic characterization of the outflows \citep{Beh03,Krae05,Mao19,Gu23,Li23}.

The combination of \xmm and \xrism observations offers unprecedented opportunities for comprehensive full-band X-ray modeling of NGC 3783 outflows. \xmm Reflection Grating Spectrometers (RGS) \citep{denH01} provide sensitivity to lower-ionization absorption features in soft X-rays. The \xrism/Resolve microcalorimeter \citep{Kelley25,Ishisaki25} delivers superior energy resolution ($\sim$5 eV FWHM at 6 keV) compared to previous missions, enabling detailed characterization of the highly ionized outflows in this active galactic nucleus. The synergy between these observatories allows for simultaneous modeling across the full X-ray band, providing complete coverage of absorption features from multiple ionization phases and enabling robust determination of physical parameters.

The spectral signature of M-shell iron line absorption in UTA is among the deepest yet observed in any AGN spectrum \citep{Sako01}. UTA features in NGC 3783 represent a particularly powerful diagnostic tool to understand the physical conditions within AGN outflows: the electron temperature, density, and ionization structure of the absorbing gas \citep{Beh01,Kasp02}. The detailed analysis of UTA features also can reveal information about the thermal structure of the outflow, the presence of multiple velocity components, and the degree of photoionization equilibrium within the absorbing material \citep{Netz2013}. In this work, we adopt the commonly used assumption of photoionization equilibrium for the warm absorber, as our analysis compares observations spanning a time baseline of about 24 years. The time variability of UTA features in NGC 3783 provides additional constraints on the geometry and physical conditions of the absorbing medium. Significant variations in the Fe UTA have been observed over timescales of weeks to months, suggesting that these features respond to changes in the ionizing continuum on relatively short timescales \citep{LiUTA}. This variability indicates that the UTA-producing regions are located at intermediate distances from the central engine, close enough to respond to continuum variations but far enough to maintain the specific ionization conditions required for M-shell iron transitions \citep{kron2003}.

This paper presents a comprehensive analysis of NGC 3783's outflows using combined \xmm and \xrism observations with particular emphasis on the comparison with the 2000--2001 observations and detailed UTA analysis. Section \ref{sect_data} describes the observational data and reduction procedures. Section \ref{sect_model} presents our full-band spectroscopic modeling approach, including the treatment of continuum components and absorption components, and the results of our spectral analysis. Section \ref{sect_discuss} discusses the difference of the outflows between 2000--2001 observation and 2024 observation. A detailed analysis of the UTA features, their variability, the kinematic structure, and the thermal stability is also discussed. Finally, Section \ref{sect_concl} summarizes our conclusions.

\section{Data Processing and Preparation}
\label{sect_data}

Our multi-wavelength campaign on NGC 3783 took place between July 17–29, 2024. Coordinated observations with \xrism, \xmm, \nustar, \chandra, Swift, \nicer, and \hst were carried out. In this paper, we analyse the combined time-averaged spectra of \xmm/RGS and \xrism/Resolve to study the multi-phase outflows. A detailed analysis of the \xrism/Resolve spectrum has been presented by \cite{Mehdi25}. \xmm observed NGC 3783 between July 20, 2024 and July 26, 2024 over 3 orbits using both the EPIC cameras \citep{Turn01,Stru01} and the Reflection Grating Spectrometer (RGS) \citep{denH01}. The \xrism observation of NGC 3783 was conducted with the gate valve closed and the open filter wheel configuration of Resolve and started on 18 July 2024 for a total Resolve duration of 451 ks (OBSID= 300050010). A summary of the observations used in this paper is given in Table~\ref{table_data}. The overall structure of the campaign and the detailed description of the data reduction and cross-calibration procedure for all the instruments are given in \cite{Kaas25}.

\begin{table}[!tbp]
\setlength{\extrarowheight}{2pt}
\caption{List of observations used in this paper (T$_\text{exp}$: Net exposure time).}
\label{table_data}
\centering
\renewcommand{\footnoterule}{}
\begin{tabular}{c c c c }
\hline \hline
Satellites  & OBSID            & Instrument	          & T$_\text{exp}$ (ks)\\
\hline
\xmm & 0923090101,      & RGS1     & 379       \\
      & 0923090201,   & RGS2     & 381    \\
      & 0923090301\\
\hline
\xrism     &  300050010    & Resolve     & 439        \\

\hline
\end{tabular}
\vspace{0.0cm}
\end{table}

\begin{figure*}[!tbp]
\centering
\hspace*{-2.2cm}\resizebox{1.22\hsize}{!}{\includegraphics[angle=0]{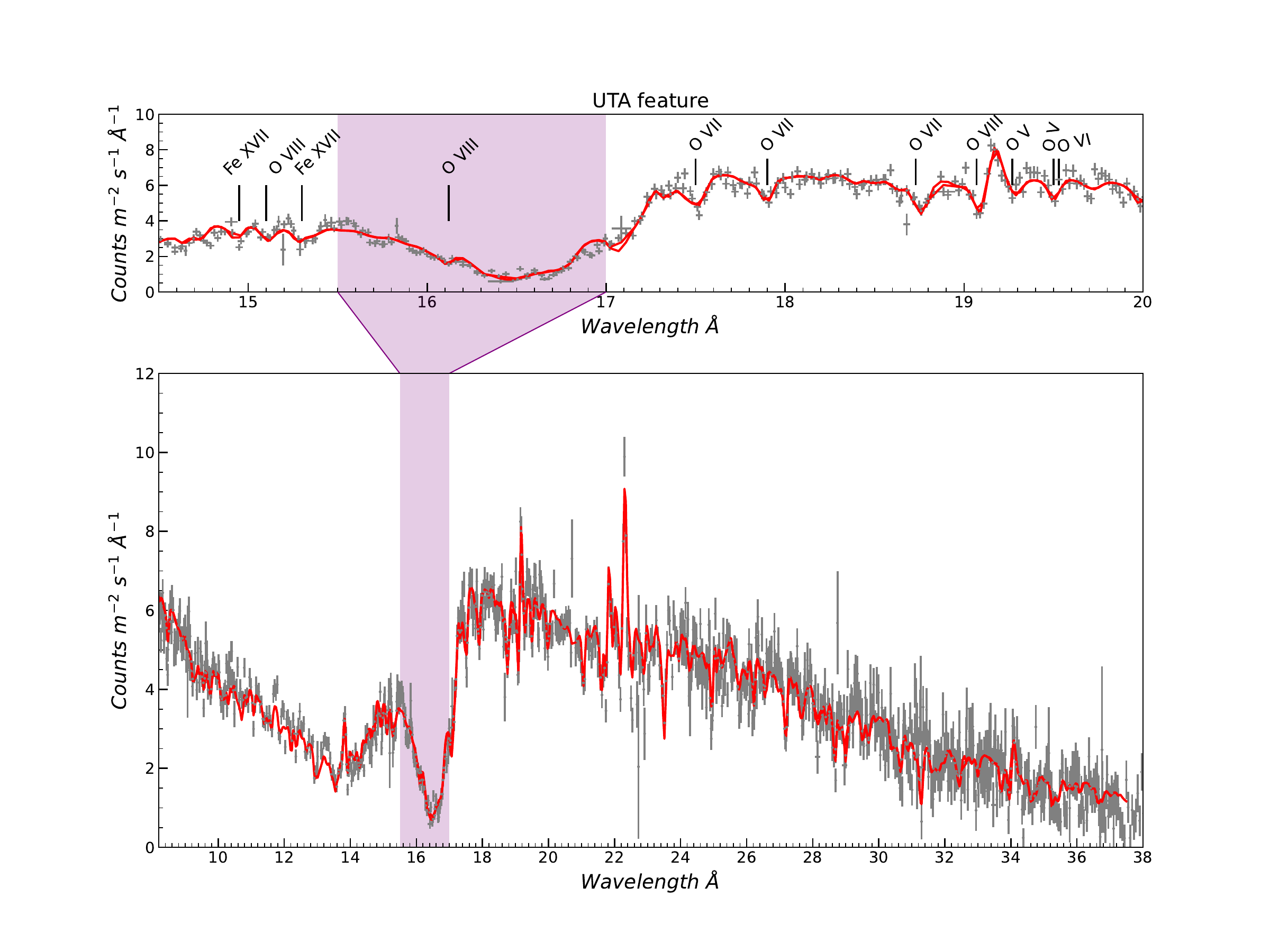}}\vspace{-1cm}
\caption{ \xmm/RGS spectrum of \ngc with our best-fit model. The top panel shows a blow-up of the spectrum near the UTA feature. The strongest absorption features are labeled. Our best-fit model (Table \ref{table_para}) is shown in red.}
\label{fig_specxmm}
\end{figure*}

\begin{figure*}[!tbp]
\centering
\hspace*{-2.2cm}\resizebox{1.22\hsize}{!}{\includegraphics[angle=0]{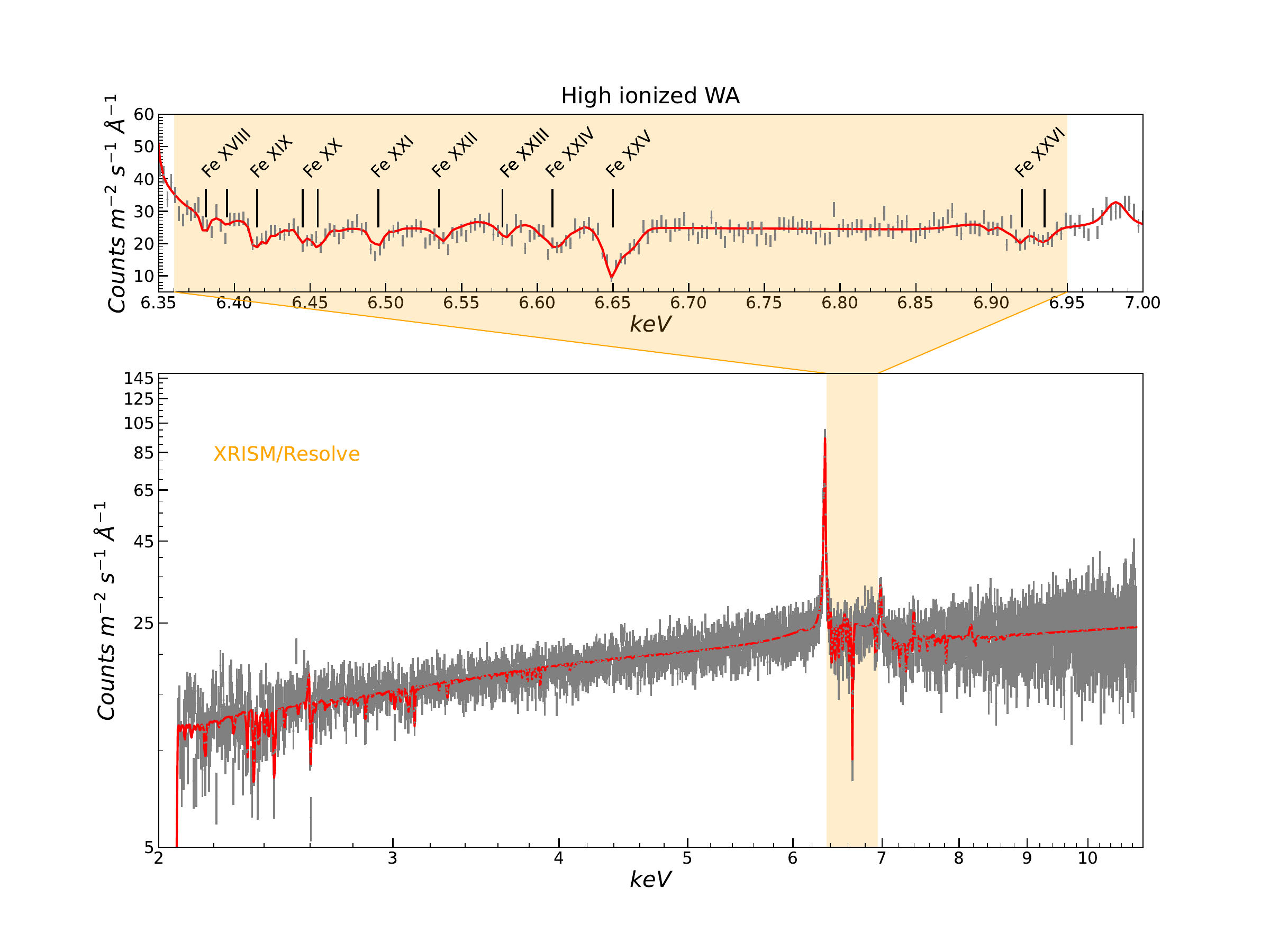}}\vspace{-1cm}
\caption{\xrism/Resolve spectrum of \ngc with our best-fit model. The top panel show the high ionization WA features. For clarity of display the spectrum in this figure is additionally binned up. The strongest emission and absorption features are labeled. Our best-fit model (Table \ref{table_para}) is shown in red.}
\label{fig_specresolve}
\end{figure*}

%====================

%###################################################
%###################################################
\section{Spectral Modeling and Analysis}
\label{sect_model}

We modeled the time-averaged \xmm/RGS and \xrism/Resolve spectra (Fig. \ref{fig_specxmm} and Fig. \ref{fig_specresolve}) simultaneously using \spex\ {\tt v3.08.02} \citep{Kaas96,Kaas24} and its latest atomic database. The spectra were fitted over the 8.3--37.5 $\mathrm{\AA}$ range for \xmm/RGS and 2.06--10.88 keV range for \xrism. The spectra were optimally binned using the {\tt rbin} task in \spex. Spectral fitting was performed with C-statistics. The cosmological redshift was fixed at 0.009730 \citep{Theu98}. Galactic X-ray absorption was accounted for using the {\tt hot} model in \spex, with a column density of $\NH = 9.59 \times 10^{20}$~cm$^{-2}$ \citep{Murp96}. The abundances of all components were set to the proto-solar values of \citet{Lod09}.

The intrinsic SED of NGC 3783 is modeled with a warm Comptonization component ({\tt comt} in \spex) to account for the soft X-ray excess, and a power-law component ({\tt pow}) representing emission from the hot corona in the hard X-ray band following \cite{Mehdi25}. The best-fit continuum parameters are listed in Table~\ref{table_contiuum}. Broad and narrow emission features originating from the X-ray photoionized emitter are reasonably fitted with the \pion model. During the fitting process, we found that the emission features in the \xmm/RGS spectrum did not vary relative to the 2000--2001 unobscured state. Therefore, the \pion emitter parameters were fixed to the values established for that epoch by \citet{Mao19}. Emission lines detected in the \xrism/Resolve spectrum were modeled with a delta function ({\tt delta} in \spex) convolved with a Gaussian profile ({\tt vgau}). Highly ionized outflows and the detailed emission lines observed with Resolve are described in \citet{Mehdi25}, while the broad \FeKa emission is discussed in \citet{Li25}. For consistency, we adopted the relativistic emission profile reported in \citet{Li25}, along with Gaussian models of the BLR and NLR components from \citet{Mehdi25}.

We modeled the absorption features of the warm absorbers with the photoionization plasma model \pion in \spex. This model computes thermal equilibrium, ionization balance, and ionic level populations from the input continuum and generates the corresponding absorption (or emission) spectrum \citep{Meh16b, Mill15}. Our observations are consistent with the intrinsic UV and X-ray continuum of the 2001 unobscured SED of NGC 3783 from \citet{Mehdi25}. Consequently, for the photoionization calculations, we adopted the continuum model from \citet{Mehdi25} which also includes a model for the optical and UV bands. Using the program {\tt xabsinput} in \spex, which runs the \pion model, we generate tables of ionic concentrations as a function of the ionization parameter ($\xi$). These tables were then used by the \xabs model in \spex to compute the absorption spectrum. This speeds up the spectral fitting process. The spatial configuration of the continuum, emitter, and warm absorbers, as well as the way these components are combined in \spex, follows the methodology applied to NGC 3783 by \citet{Mao19} and \citet{Mehd17,Mehdi25}. We first adopt the five warm absorber (WA) components from \cite{Mehdi25}. These components match the \xrism/Resolve spectra well, but not when combined with the \xmm/RGS spectra (C-stat $\sim$ 17966/4300). We then add components C2, B2, and A4 step by step, resulting in C-stat improvements of $\Delta C \sim 6166$, 1124, and 5079, respectively. The ionization parameter ($\xi$), column density (\NH), outflow velocity (\vout), and turbulent velocity (\sigv) of the \xabs components were fitted, assuming full covering fractions for all components.

We identify a total of eight \xabs absorption components associated with the warm absorber. Components are labeled alphabetically in ascending order of outflow velocity (\vout). For components with comparable velocities, numerical subscripts are used in descending order of ionization parameter ($\xi$). The best-fit models are shown in Fig.~\ref{fig_specxmm} and Fig.~\ref{fig_specresolve}, while the best-fit parameters of the warm absorber components in the \xmm/RGS and \xrism/Resolve time-averaged spectra are listed in Table~\ref{table_para}. The best-fit C-statistic ratio relative to the expected value is 5597/4300, corresponding to a reduced C-statistic of $\sim$1.30.

The highest ionization components (Comps. A1, B1 and C1) reproduce the \fexxvi and \fexxv absorption at different velocities. Comps. A2, B1-2 and C2 account for absorption from \ion{Fe}{xvii}–\ion{Fe}{xxiv}, as well as the \ion{Si}{xiv}, \ion{Si}{xiii}, \ion{S}{xiv}, \ion{S}{xv}, \ion{Mg}{xii} and \ion{O}{viii} transitions. Comp. A4 produces most of the oxygen absorption lines from \ion{O}{vii} down to \ion{O}{v}. The low ionization transitions around 31--37.5 $\mathrm{\AA}$ are difficult to constrain due to decreasing data quality at lower energies. The majority of the UTA feature (\ion{Fe}{viii}–\ion{Fe}{xiii}) is produced by Comp. B3, which also has the highest column density and contributes significantly to absorption by \ion{C}{v}, \ion{C}{vi}, \ion{N}{vi}, \ion{N}{vii}, \ion{O}{vii}, \ion{O}{viii}, and \ion{Ne}{viii}.

The \xmm/RGS spectrum also provides sensitivity to potential Galactic or intergalactic absorption at zero redshift. \ion{O}{VII} (21.60 $\mathrm{\AA}$ in rest frame) was previously detected by \cite{Gupta2012} in their analysis of Chandra archival data from NGC 3783 to study the warm-hot gas of the circumgalactic medium (CGM) of the Milky Way. We applied the \hot model in \spex to represent the CGM gas. Our results show that the temperature and the column density of the gas are $0.12\pm0.08$ keV and $(5.2\pm0.9)\times$10$^{19}$ cm$^{-2}$, respectively. This result is consistent with \cite{McClain2024}.

\begin{table}[!tbp]
\setlength{\extrarowheight}{2pt}
\caption{Best-fit parameters of the continuum of
NGC 3783.}
\label{table_contiuum}
\centering
\renewcommand{\footnoterule}{}
\begin{tabular}{c l l r }
\hline \hline
Model  & Parameter            & Values	          & Units\\
\hline
\tt comt & Norm      & 4.0$\pm0.7$     & 10$^\text{56}$ ph s$^{-1}$ keV$^{-1}$      \\
       & T$_\text{c}$    & 0.28$\pm0.03$     & keV    \\
       & $\tau$     & 11.0$\pm1.0$     &     \\
\hline
\tt pow     &  Norm    &  4.34$\pm0.01$   & 10$^\text{51}$ ph s$^{-1}$ keV$^{-1}$      \\
            &  $\Gamma$    &1.80$\pm0.01$     &         \\

\hline
\end{tabular}
\vspace{0.0cm}
\end{table}

Both this work and \citet{Mehdi25} adopt the same unobscured SED. While \citet{Mehdi25} focused on highly ionized warm absorbers, our joint \xmm and \xrism analysis extends to lower and intermediate ionization states. The ionization components identified here (A1, A2, B1, B3, C1) correspond to those in \citet{Mehdi25} (A1, A2, B, A3, C, respectively, see Table~\ref{table_para} Comp$^*$). In addition, we identify several intermediate-ionization components (B2, C2) and lower-ionization components (A4) that account for the features absent in \citet{Mehdi25}. We also include the broad, sub-relativistic absorption component Comp. X in \citet{Mehdi25} for consistency.

\begin{figure}
\centering
\hspace*{0.0cm}\resizebox{1\hsize}{!}{\includegraphics[angle=0]{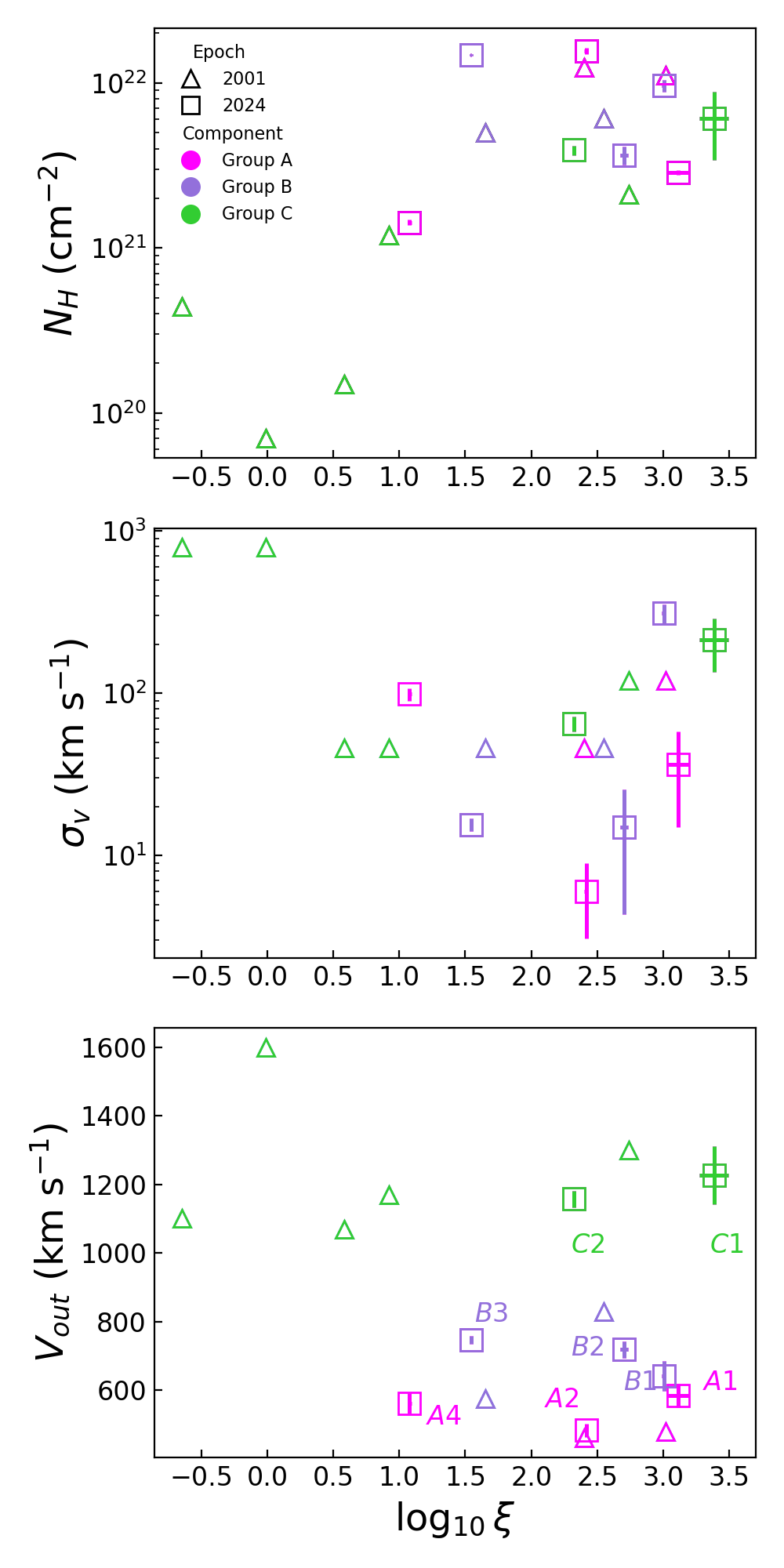}}\vspace{0.0cm}
\vspace{0.0cm}
\caption{Relations between the parameters of the eight outflow components derived from the \xrism Resolve spectrum of \ngc. The top panel displays the column density \NH as functions of the ionization parameter (\logxi). The middle and bottom panels show the turbulent velocity (\sigv) and outflow velocity (\vout) as functions of \logxi, respectively. The component label (Table \ref{table_para}) for each data point is shown along the lower edge of the bottom panel. The triangle and square represent the WAs in 2000--2001 \citep{Mao19} and 2024 observations, respectively. The different colour represents the components grouped by \vout.}
\label{fig_rel}
\end{figure}

\begin{table}[!tbp]
\begin{minipage}[t]{\hsize}
\setlength{\extrarowheight}{2pt}
\setlength{\tabcolsep}{3pt}
\caption{Best-fit parameters of the outflow components that we have determined from the \xmm RGS and \xrism Resolve spectra of \ngc. 
}
\label{table_para}
\centering
\small
\renewcommand{\footnoterule}{}
\begin{tabular}{c | c c c c c}
\hline \hline
Comp.  & \vout              & \logxi	          & \NH	           	  & \sigv   &  Comp$^*$  \\
       & (\kms)             & (erg~cm~s$^{-1}$)   & ($10^{21}$~\cm)   & (\kms)  &   \\
\hline
A1      & $583 \pm 35$      & $3.12 \pm 0.08$     & $2.9 \pm 0.1$       & $<36$ & A1\\
A2      & $482 \pm 19$      & $2.42 \pm 0.01$     & $15.6 \pm 0.7$       &$6 \pm 3$ & A2 \\
A4       & $560 \pm 30$      & $1.08 \pm 0.02$     & $1.43\pm 0.05$       & $99 \pm 9$  \\
B1      & $641 \pm 44$      & $3.00 \pm 0.01$     & $ 9.7\pm 0.8$    & $312 \pm 41$ & B\\
B2     &  $718 \pm 25$  & $2.70 \pm 0.03$     & $3.7 \pm 0.5$        & $15 \pm 11$   \\
B3     & $747 \pm 12$    & $1.54\pm 0.01$     & $14.8\pm 0.2$        & $16 \pm 1$  & A3 \\
C1       & $1226 \pm 84$  & $3.38 \pm 0.11$   & $7.1\pm 2.7$     & $232\pm 73$  & C \\
C2      & $1157 \pm 23$  & $2.32 \pm 0.01$        & $3.9 \pm 0.3$     & $65 \pm 7$   \\
\hline
\multicolumn{5}{c}{C-stat / expected C-stat = 5597 / 4300 = 1.30} \\
\multicolumn{5}{c}{C-stat / degrees of freedom = 5597 / 4289 = 1.30} \\
\hline
\end{tabular}
\end{minipage}
\tablefoot{
Components are labeled with a letter in ascending order of outflow velocity \vout, and those with similar \vout are further sub-labeled by a number in descending order of ionization parameter $\xi$. Comp$^*$ represents the corresponding component name in \cite{Mehdi25}.}
\vspace{0.0cm}
\end{table}

The best-fit models are shown in Fig.~\ref{fig_specxmm} and Fig.~\ref{fig_specresolve}, with parameters listed in Table~\ref{table_para}. Reported uncertainties correspond to the 1$\sigma$ confidence level. In Fig.~\ref{fig_rel}, we examine correlations among the absorber parameters. Thermal stability curves (S-curves) for the warm absorbers are presented in Fig.~\ref{fig_scurve}. Finally, in Fig.~\ref{fig_trans}, we compare the UTA component between the 2000--2001 and 2024 observations, showing the spectra and best-fit models (upper panel), transmission curves (lower panel), and ionic column density distributions for Fe and O (Fig.~\ref{fig_ionic}). These results are discussed in detail in the following section.

\section{Discussion}
\label{sect_discuss}
\subsection{Comparison with the 2001 unobscured epoch}

In this paper, we jointly fit the \xmm/RGS and \xrism/Resolve spectra of \ngc. Our spectroscopic and photoionization modeling identifies eight distinct absorption components, spanning a wide range of ionization states ($\logxi = 1.08$ to 3.38) and outflow velocities (480--1230 km/s).

Previous X-ray studies of \ngc\ during its historical unobscured epoch revealed nine distinct ionization and velocity components. The 2000--2001 observations \citep{Mao19}, based on \chandra and \xmm data, showed a larger number of low-ionization absorbers ($\logxi = -0.65$ to 0.92). By contrast, our 2024 analysis reveals a greater number of medium- and high-ionization components. As shown in Fig.~\ref{fig_rel}, the triangles and squares represent the 2000--2001 and 2024 WA components, respectively.

The column densities and turbulent velocities remain broadly consistent between the two epochs, but the earlier data show more high-velocity components at relatively low ionization. This difference may reflect changes in our line of sight through a clumpy, stratified wind \citep{Blu05}, with certain low-ionization, fast-moving clouds no longer intersecting the sight-line in 2024 \citep{Mehd17}. Alternatively, changes in the accretion conditions may have altered the balance of ionization and acceleration, reducing the presence of low-ionization, high-velocity gas \citep{Noda18}. The possibility that some components are transient, appearing or dissipating over decade-long timescales, cannot be ruled out \citep{kron2007}.

The Absorption Measure Distribution (AMD)—defined as the differential column density as a function of ionization parameter \citep{Holc07, Beh09, Ster14, Adhi19}—provides a powerful diagnostic of the physical conditions in ionized outflows and insight into their launching and driving mechanisms. Our results show that $N_{\text{H}}$ rises from $\logxi=1.08$ to $\logxi=1.54$, followed by a relatively flat AMD shape across the medium- and high-ionization regime ($\logxi < 3.38$; Fig.~\ref{fig_rel}, top panel). The higher ionized outflows are also faster, with both higher outflow velocities and larger turbulence. Compared to \cite{Mehdi25}, our analysis includes more low- and intermediate-ionization components, but the AMD in the overlapping ionization range shows the same general trend. These AMD and velocity patterns are complex and cannot be attributed to a single, simple wind-driving mechanism.

The column density of the UTA-dominant component in 2024 (Comp. B3) has increased by a factor of three relative to its 2000--2001 counterpart, while its ionization parameter, turbulent velocity, and outflow velocity remain largely unchanged. Thanks to the high-resolution \xrism/Resolve spectrum, we are also able to characterize highly ionized outflows in unprecedented detail, identifying new components (Comp. C1--2) that were absent in 2001. The turbulent velocities (middle panel, Fig.~\ref{fig_rel}) and outflow velocities (bottom panel, Fig.~\ref{fig_rel}) cover broadly similar ranges in 2001 and 2024. A more detailed discussion of the origin of the warm absorbers, and the variation of Comp. B3 over 24 years, is presented below.

\subsection{Radius of the warm absorbers}
The radius $r$ corresponding to the location of the outflow provides important clues about the origin and acceleration mechanisms of warm absorbers (WAs). We first consider the minimum launch radius, \rmin, which can be estimated by equating the outflow velocity to the local escape velocity:
   \begin{equation}
      \rmin= \frac{2G\MBH}{v_{\text{out}}^{\text{2}}} \,,
   \end{equation}
where the black hole mass $\MBH = 2.82\times10^7 M_{\odot}$ is from \cite{Bentz2021}. We assume that the thickness of the WA ($\Delta r$) does not exceed the radius at which it is launched, $V_{\text{f}}$ is the volume filling factor, and 
$\Delta r=\NH/n_{\text{H}}V_\text{f}$. According to the definition of the ionization parameter $\xi=\lion/n_{\text{H}} r^{2}$, imposing $\Delta r = r$ yields an upper limit of the radius:
   \begin{equation}
     \rmax= \frac{L_{\text{ion}}V{_\text{f}}}{\xi N{_\text{H}}} \,,
   \end{equation}
where the 1-1000 Ryd ionizing luminosity is $L_{\text{ion}}=6.4\times10^{43}$ \ergs \citep{Mehd17}. Since the volume filling factor may differ for each component, we adopt the volume filling factor as 1 for a homogeneous and volumetrically thick outflow. Assuming that the X-ray emitting region has a size no larger than 20 gravitational radii, and treating the transverse velocity of the WA as Keplerian velocity $v_{\text{kepler}}=(G\MBH/r)^{\frac{1}{2}}$, we can derive the timescale of transverse motion of a WA that moves across the line of sight (\citealt{Li23}):
   \begin{equation}
   \label{crossingtime}
      t_{\text{cross}}=\frac{20}{c^2}(G\MBH)^{\frac{1}{2}}(\frac{L_{\text{ion}}}{\xi})^{\frac{1}{4}}n{_\text{H}}^{-\frac{1}{4}},
   \end{equation}
The sound speed crossing time through the thickness of the WA is the minimum timescale over which the WA can adjust its internal pressure balance across its thickness. So we calculate the sound speed crossing time $t_{\text{th}}=\Delta r/v_{\text{s}}$ to estimate the thermal adjustment. $v_{\text{s}}$ is the sound speed that can be computed from the temperature of the photoionized gas 
as provided by the \pion model in \spex. Another relevant time scale is the cooling timescale, i.e. the time required for the gas to restore thermal balance following a perturbation in the temperature. It can be estimated as:
   \begin{equation}
   \label{crossingtime}
      t_{\text{cooling}}=2.30\times\frac{3/2n_{\text{H}}kT}{\Lambda(T)},
   \end{equation}
where the numerator is the total thermal energy density and the denominator the net cooling rate in the gas per unit volume, with $\Lambda (T)$ representing the total cooling rates, which are also obtained from the \pion calculations.

Using the best-fit results, we summarize all derived quantities, including WA distance, thickness, transverse crossing time, sound speed, dynamical timescale, and cooling time in Table~\ref{table_estvalue}. These estimates of $\Delta r$, $t_{\text{cross}}$, $t_{\text{th}}$, and $t_{\text{cooling}}$ are based on the $r_{\text{min}}$ values, since the $r_{\text{max}}$ values approach infinity and provide weaker constraints.

\begin{table*}[h!]
\centering
\caption{The derived parameters of WAs.}
\label{table_estvalue}
\renewcommand{\arraystretch}{1.2}
\begin{tabular}{c c c c c c c c c c c}
\hline\hline
WAs&$r_{\min}$ & $r_{\max}$ & $\Delta r_{\min}$ & $t_{\rm cross,max}$ & $v_{\rm s}$ &$T$ & $\Lambda(T)$ & $t_{\rm th}$& $t_{\rm cooling}$ \\  
 & & &$\sim r^{2}$& $\sim r$& & & &$\sim r^{2}$ &$\sim r^{2}$ \\
&(pc) & (pc) & (pc) & (yr) & (km/s) & (keV) & ($10^{-14}$ W/m$^{3}$) & (yr)  & (yr)\\
\hline
A1&0.71 & 2 & 0.09  & 0.06 & 353 &0.49 &0.01 & 253 & 735\\
A2&1.04 & 5 & 0.21 & 0.08 & 112 &0.05 &0.05 & 1866 & 40 \\
A4&0.77 & 1213 & 0.49$\times10^{-3}$ & 0.07 & 28  &0.004 &90& 17& 0.07\\
B1&0.59 & 2 & 0.16 & 0.06 & 258 &0.26 &0.04 & 621&216 \\
B2&0.47 & 11 & 0.02 & 0.05 & 151 &0.09 &0.37 & 127& 25 \\
B3&0.43 & 40 & 4.70$\times10^{-3}$ & 0.05 & 42 &0.007 &135 & 111&0.09\\
C1&0.16 & 1 & 0.02  & 0.03 & 586 &1.33 &2 & 31 & 125 \\
C2&0.18 & 25 & 1.30$\times10^{-3}$ & 0.03 & 89 &0.03 &93 & 14 & 0.57\\
\hline
\end{tabular}
\tablefoot{$r_{\text{min}}$ and $r_{\text{max}}$ are the minimum and maximum distance, respectively. $\Delta r_{\text{min}}$ is the minimum thickness of WAs. $t_{\rm cross,max}$ represents the transverse motion time that WA crossing our line of sight. $t_{\text{cooling}}$ is the cooling time for every WA. $v_{\text{s}}$ is the sound speed. T is the electron temperature. $t_{\text{th}}$ is the sound speed crossing time. $\Lambda(T)$ is the total cooling rate. $v_{\text{s}}$, $T$ and $\Lambda(T)$ are all obtained from the \pion calculations.}
\end{table*}

Warm absorbers (WAs) are known to vary on both short and long timescales. Short-term variability (days–weeks) is typically driven by changes in the ionizing flux, which alter the ionization balance while leaving the total column density largely unchanged. In contrast, long-term variability is more likely associated with thermal instabilities or bulk transverse motions of the absorbing gas. Indeed, although the overall shape of the X-ray absorption spectrum of NGC 3783 has remained stable since 1996 \citep{Kasp02}, the differences observed between 2000--2001 and 2024 are best explained by thermal or dynamical processes, or a combination of the two.

\begin{figure}
\centering
\hspace*{0.0cm}\resizebox{1\hsize}{!}{\includegraphics[angle=0]{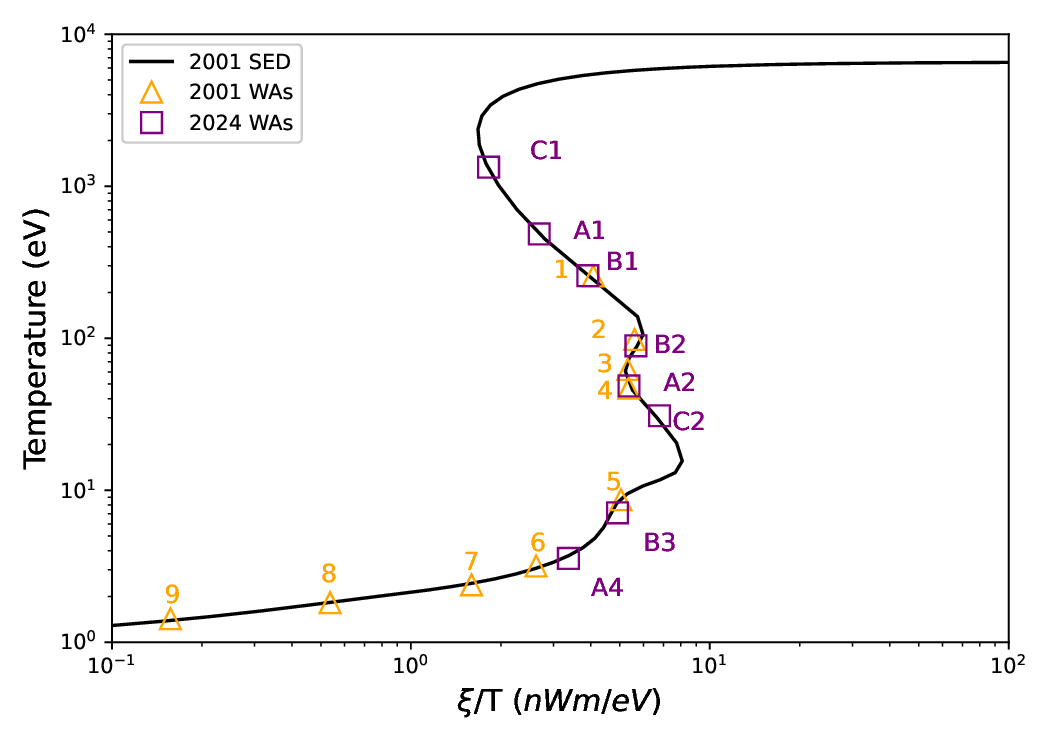}}\vspace{0.0cm}
\vspace{0.0cm}
\caption{Curve of thermal stability. The SED is from \cite{Mehd17} in the 2001 unobscured state. The orange triangle and purple square corresponds to the 2000--2001 and 2024 WAs, respectively. The name of the WAs in 2000--2001 and 2004 are marked by the same colour as the data points.}
\label{fig_scurve}
\end{figure}

Thermal stability can be assessed using the S-curve in the $(T, \xi/T)$ plane, where the equilibrium temperature depends on the ionizing SED. We computed stability curves with the \pion\ model using the 2001 SED (Fig.~\ref{fig_scurve}). The 2001 WAs (orange triangles) and the 2024 WAs (purple squares) occupy broadly similar regions, with low-ionization components lying on the stable branch (Comps. 5--9 in 2001; Comps. B2--3, A4 in 2024). Comp. C1 and A1, however, reside on the unstable branch with the highest temperature, while Comps. A2, A3, and B1 overlap with the 2000--2001 Comps. 1--4, also near the unstable branch. Components in this regime are expected to respond rapidly to modest changes in flux or density. The total column density of the 2024 WAs ($N_{\text{H}} = 5.52 \times 10^{26}$ m$^{-2}$) exceeds that of 2001 ($N_{\text{H}} = 3.86 \times 10^{26}$ m$^{-2}$), implying the addition of new material. However, the summed column density of the intermediate-ionization components is nearly identical between the two epochs (Comp. 1--4 in 2000--2001: $3.17 \times 10^{26}$ m$^{-2}$; Comp. A2, A3, B1, C2 in 2024: $3.28 \times 10^{26}$ m$^{-2}$), and their comparable ionization parameters, velocities, and turbulence suggest that they may represent the same gas, reconfigured over time. By contrast, Comp. A1, C1 have no analogue in the 2001 data and likely represent newly emerged absorbers along our line of sight, or undetectable in the Fe K band with the lower resolution CCD instruments in that band used in 2001.

Finally, the transverse crossing timescales of all WAs are shorter than 24 years (Table~\ref{table_estvalue}), indicating that the material observed in 2024 does not correspond to the exact same portions of the WA structures seen in 2001. Instead, we are likely probing different segments of the same global outflow, supplemented in part by newly emerged components.

\subsection{The variation of the UTA component}
\begin{figure}[!tbp]
\centering
\hspace*{-0.0cm}\resizebox{1.00\hsize}{!}{\includegraphics[angle=0]{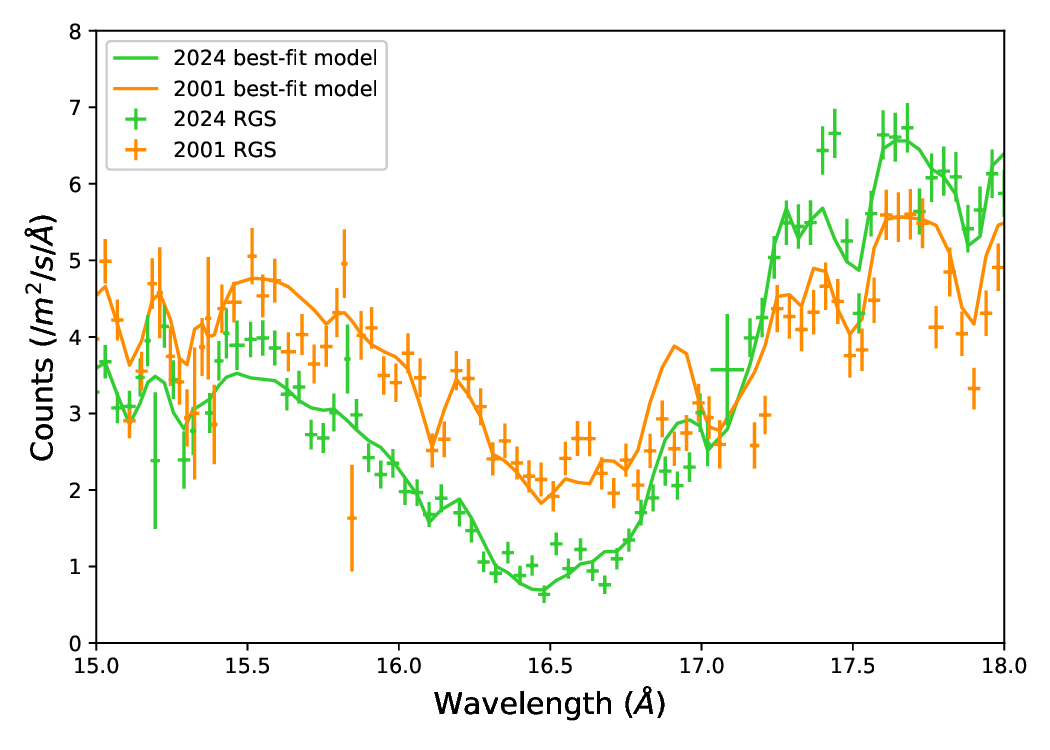}}\vspace{-0.0cm}
\hspace*{-0.0cm}\resizebox{1.00\hsize}{!}{\includegraphics[angle=0]{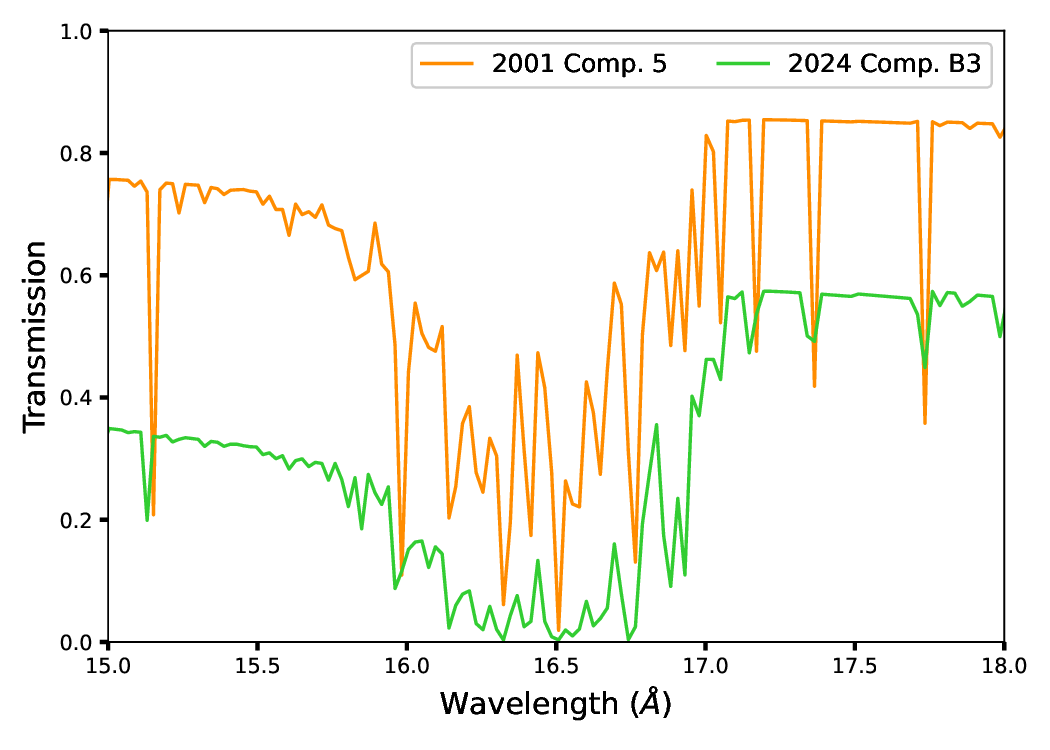}}
\caption{Upper panel: The RGS spectra of 2001 and 2024 with their best-fit model in the UTA energy band. Lower panel: Transmission model of the UTA in 2001 (Comp. 5) and 2024 (Comp. B3).}
\label{fig_trans}
\end{figure}

The Fe M-shell 2p--3d unresolved transition array (UTA) between 16 and 17 $\mathrm{\AA}$, together with the \ion{O}{vii} and \ion{O}{viii} photoelectric bound-free K-edges (14.2 and 16.8$\mathrm{\AA}$), constitutes the most significant broad absorption features in NGC 3783. These features allow us to readily identify the dominant component in the 2024 observations corresponding to that seen in 2000--2001. The feature has deepened considerably over the past 24 years (Fig.~\ref{fig_trans}, upper panel). In Fig.~\ref{fig_trans} (lower panel), we compare the transmission of the UTA component in 2000--2001 (Comp. 5; \citealt{Mao19}) with that in 2024 (Comp. B3). The 2024 transmission shows stronger saturation than Comp. 5 in 2001, although not all of the lines in Comp. B3 are saturated.

Over the 24 years, the ionization parameter has remained nearly constant ($\logxi=1.65$ in 2001 and $\logxi=1.54$ in 2024), while the column density increased by a factor of three, from $N_{\text{H}} = 5.00 \times 10^{21}$cm$^{-2}$ to $N_{\text{H}}=14.83\times10^{21}$ cm${^\text{-2}}$. The outflow velocity rose to 747 km/s, while the turbulent velocity decreased to 16 km/s. Because Comp. B3 is partially saturated, the derived column density should be regarded as a lower limit.

\begin{figure}[!tbp]
\centering
\hspace*{-0.3cm}\resizebox{0.92\hsize}{!}{\includegraphics[angle=0]{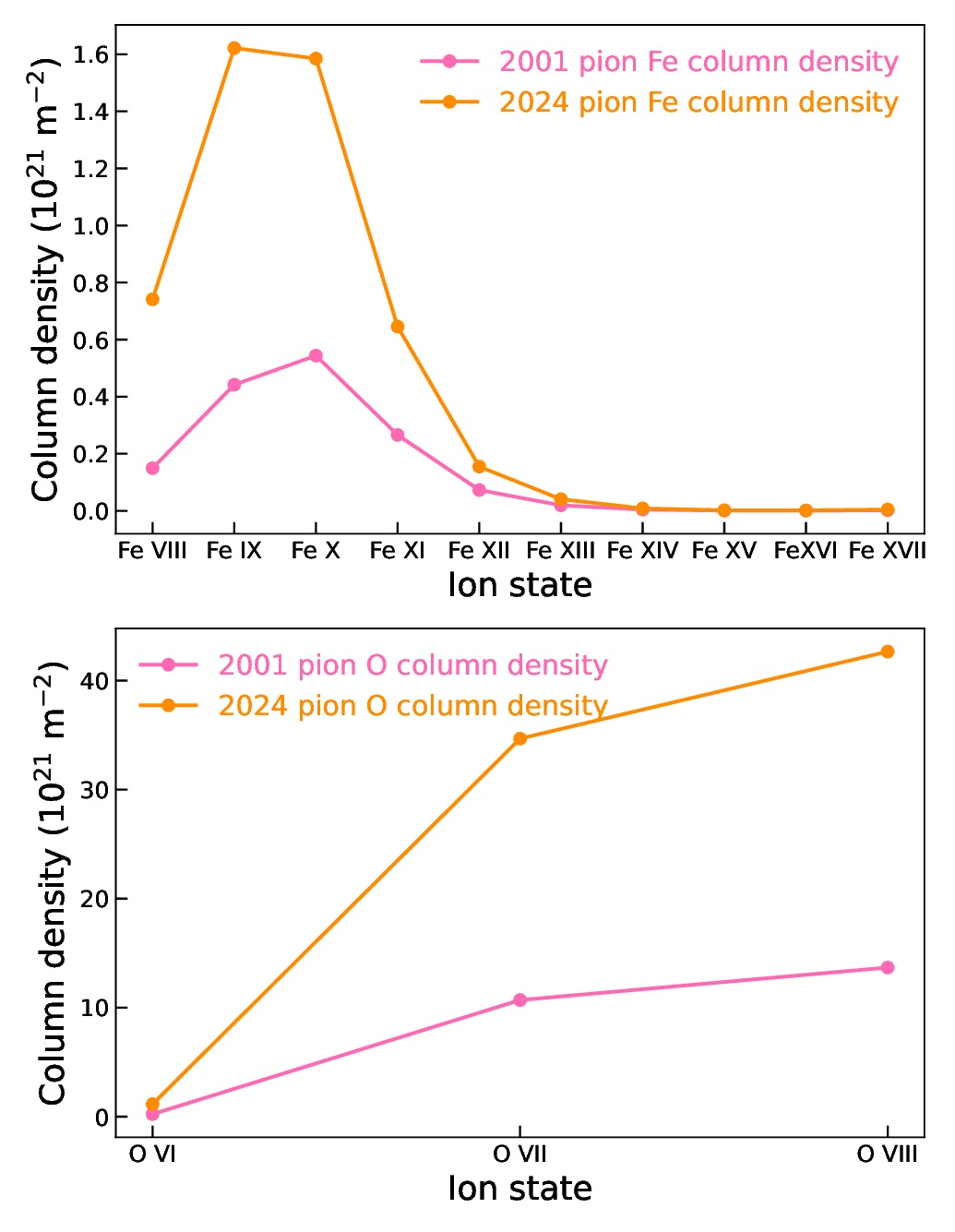}}\vspace{-0.0cm}
\caption{Column densities of Fe and O in the UTA component B3. The upper panel shows the column density of \ion{Fe}{viii} to \ion{Fe}{XVII} calculated by \pion (green) and \slab (orange) model. The lower panel shows the column density of \ion{O}{vi} to \ion{O}{VIII} calculated by \pion (green) and \slab (orange) model.}
\label{fig_ionic}
\end{figure}

We present the ionic column densities of iron (\ion{Fe}{viii}--\ion{Fe}{xvii}) and oxygen (\ion{O}{vi}--\ion{O}{viii}) for the 2000--2001 and 2024 UTA components in Fig.~\ref{fig_ionic}. These values are obtained from the \pion models using the best-fit results for both epochs. The ionic column densities of each element are determined self-consistently within the \pion model in \spex \citep{Meh16b}, which solves the photoionization and energy balance equations using the SED of the source as constructed within the fit. For a given N$_{\text{H}}$ and \logxi, \pion computes the ionization balance across all ionic species and uses the resulting ionic fractions together with the assumed elemental abundances to obtain the ionic column densities. Significant deviations from the 2001 curves are evident for \ion{Fe}{viii}--\ion{Fe}{xii} and \ion{O}{vii}--\ion{O}{viii}, consistent with the threefold increase in column density. The peak near \ion{Fe}{IX} and the rising trend from \ion{O}{vi} to \ion{O}{viii} remain essentially unchanged, reflecting the stability of the ionization parameter.

The spectral–timing method, which analyses plasma responses to variability in the ionizing luminosity, provides tighter constraints on the density and distance of the WAs \citep{Sad25}. Assuming that Comp. B3 in 2024 is the same physical component as Comp. 5 in 2000--2001, we adopt the density limit of $n_{\text{e,B3}} < 10^{12.3}$ m$^{-3}$ and a distance $r_{\text{B3}} < 0.27$ pc from \cite{LiUTA}. From these values, we derive a transverse crossing time of $t_{\text{cross,B3}} \sim 0.04$ yr (Eq.\ref{crossingtime}). As shown by the S-curve in Fig.\ref{fig_scurve}, the temperature and pressure of Comp. B3 are both lower than those of Comp. 5, while remaining on the stable branch. Several scenarios could explain the observed column density variation of Comp. B3 over the 24 years:

\paragraph{Same absorber, intrinsic variation.} 
Although 24~years have passed, the expected radial displacement is only \(\sim 0.01\ \mathrm{pc}\), implying that it is plausible we are still observing the same WA. If the same escaping wind was present in both 2000--2001 and 2024, the factor-of-three increase in column density would require changes in either density or physical thickness. However, the thermal timescale (\(t_{\text{th}} \sim 111\)~yr; Table~\ref{table_estvalue}) is much longer than the 24-year interval, making global compression or geometric thickening unlikely. The very short transverse crossing time (\(t_{\text{cross}} \sim 0.05\)~yr) instead suggests a transient line-of-sight effect, such as overlapping clumps or azimuthal inhomogeneities produced by orbital shear.

Archival \textit{Chandra}/HETGS variability studies from 2001 also suggest that this component could be a failed-wind candidate \citep{fwind2004,LiUTA}. In this scenario, the increase in outflow velocity from 575 to 747~km\,s\(^{-1}\) implies that the failed wind is observed during its rising phase.

We model the motion by assuming the failed wind is launched radially outward from radius \(r_0\) with initial velocity \(v_0\), moving in a Newtonian gravitational potential. Pressure forces, radiation driving, magnetic fields, and general relativistic corrections are negligible at these distances. The turning radius is
\begin{equation}
r_{\text{turn}}=
\frac{1}{
\displaystyle \frac{1}{r_0} - \frac{v_0^2}{2GM}
}.
\end{equation}
The time required to reach the turning point is
\begin{equation}
t_{\rm turn}
=
\int_{r_0}^{r_{\text{turn}}}
\frac{dr}{v(r)},
\qquad
v(r)
=
\sqrt{
v_0^2 + 2GM\left(\frac{1}{r}-\frac{1}{r_0}\right)
}.
\end{equation}
The subsequent fall-back time is simply $t_{\text{fall}}$ = 2 $t_{\text{top}}$. We adopt the observed velocity of 747 km/s as $v_0$. Using this framework, three possible cases emerge: \textbf{(A)} Adopting \(n_{\text{e,B3}} < 10^{12.3}\ \mathrm{m^{-3}}\) and \(r_{\text{B3}} < 0.27\ \mathrm{pc}\), we obtain a turning time of \(1588\)~yr and turning radius of \(<0.69\)~pc. \textbf{(B)} If the wind is currently at its turning point after 24~years, the launching radius is \(r_0 = 0.06\)~pc, with a density of \(n_e = 5.90\times 10^{13}\ \mathrm{m^{-3}}\). \textbf{(C)} If the 24-year interval corresponds to the full fall-back time, then \(r_0 = 0.04\)~pc with \(n_e = 1.13\times 10^{14}\ \mathrm{m^{-3}}\).

During 24~years, Comp.~B3 also undergoes significant azimuthal rotation due to Keplerian motion. For the three cases above, the accumulated rotation angles are approximately
$3^\circ$, $26^\circ$, and $43^\circ$, corresponding to transverse displacements of $1.2\times 10^4$, 2.3$\times 10^4$, and 2.6$\times 10^4$ $r_{\text{g}}$. These distances greatly exceed the typical size of the X-ray corona ($\sim$ 20 $r_{\text{g}}$), implying that the material currently in our line of sight is no longer the same part observed 24~years earlier.

In summary, in this scenario Comp. B3 is likely a filamentary, clumpy structure with short internal crossing times. The observed 24-year variation in column density could naturally arise from transverse motion across our line of sight, causing different substructures to be sampled at different epochs.

\paragraph{Replenishment by new material.} Alternatively, Comp. 5 may have accreted new material, evolving into the Comp. B3 seen in 2024. If the column density increase were due to condensation from hotter components, the unstable branch in Fig.~\ref{fig_scurve} would need to provide sufficient additional material. However, the total column density available there is insufficient to account for the observed growth. Moreover, the replenished material should share the kinematic properties of its source, implying higher turbulent velocities than observed in Comp. B3 (\sigv > 16 km/s). For this scenario to hold, the newly added material must have both a higher outflow velocity ($v_{\text{out}} > 747$ km/s) and unusually low turbulence (\sigv < 16 km/s). Determining the origin of such material is beyond the scope of this work.

\paragraph{A new component along the line of sight.}

Finally, Comp. B3 may be an entirely new absorber that has entered our line of sight since 2001. It shares a similar ionization parameter with Comp. 5 but differs in column density, turbulence, and outflow velocity, and is located beyond 0.43 pc (Table~\ref{table_estvalue}).

\section{Conclusions}
\label{sect_concl}

We have analysed joint time-averaged \xmm/RGS and \xrism/Resolve spectra of NGC 3783 from the 2024 observations, revealing the rich complexities of eight warm absorber (WA) components. The main results are summarized as follows:

1.The eight outflows span velocities from 461 km/s up to 1262 km/s, with ionization parameters ranging from $\logxi = 1.08$ to 3.38, and column densities between $1.43$ and $15.63 \times 10^{21}$ cm$^{-2}$.

2.The physical properties of the 2024 WAs fall within a similar overall range to those observed in 2000--2001. However, due to changes in column density and velocity, we cannot directly match individual WA components between the two epochs, except for the one dominating the UTA feature. In addition, we identify new components, C1--2, which have the highest and middle ionization states respectively. These component was not present in the 2000--2001 data and may represent WAs that have only recently entered our line of sight.

3.The column density of the component dominating the UTA feature (Comp. B3) has increased by a factor of three, its outflow velocity has risen by $\sim$172 km/s, and its turbulent velocity has decreased to 16 km/s. However, its ionization state remains comparable to that observed in 2000--2001, with slightly shift in the peak Fe ionization state. These changes indicate that Comp.~B3 is either a newly appearing WA within \(<0.43\)~pc, or the same WA observed in 2000--2001 undergoing failed-wind dynamics and transverse structural evolution.

\begin{acknowledgements}

K. Zhao acknowledges support from the Chinese Scholarship Council (CSC) and Leiden University/Leiden Observatory, as well as SRON. SRON is supported financially by NWO, the Netherlands Organization for Scientific Research. M.Signorini acknowledges support through the European Space Agency (ESA) Research Fellowship Program in Space Science.

\end{acknowledgements}

\bibliographystyle{aa}
\bibliography{references}

\end{document}